\documentstyle[12pt]{article}
\newcommand{\extraspace}{\addtolength{\abovedisplayskip}{2mm}
                        \addtolength{\belowdisplayskip}{2mm}
                        \addtolength{\abovedisplayshortskip}{2mm}
                        \addtolength{\belowdisplayshortskip}{2mm}}
\newcommand{\be}{\begin{equation}\extraspace}

\newcommand{\ee}{\end{equation}}

\newcommand{\bea}{\begin{eqnarray}\extraspace}
\newcommand{\beastar}{\begin{eqnarray*}\extraspace}
\newcommand{\eea}{\end{eqnarray}}
\newcommand{\eeastar}{\end{eqnarray*}}

\newcommand{\nonu}{\nonumber \\[2mm]}

\newcommand{\strutje}{\rule[-1mm]{0mm}{4mm}}
\newcommand{\str}{\rule[-2.5mm]{0mm}{9mm}}
\newcommand{\strr}{\rule[-2.5mm]{0mm}{6mm}}
\newcommand{\stru}{\rule[0mm]{0mm}{6mm}}

\newcommand{\half}{\frac{1}{2}}
\newcommand{\third}{\frac{1}{3}}
\newcommand{\quart}{\frac{1}{4}}

\newcommand{\eps}{\epsilon}

\newcommand{\chih}{\widehat{\chi}}

\newcommand{\eg}{{\em e.g.}}
\newcommand{\ie}{{\em i.e.}}
\newcommand{\cW}{{\cal W}}
\newcommand{\vac}{|0\rangle}

\newcommand{\ch}{{\rm ch}}
\newcommand{\chh}{\widehat{\ch}}
\newcommand{\mod}{{\rm mod}}

\newcommand{\whsun}{\widehat{su(n)}}
\newcommand{\whSUn}{\widehat{SU(n)}}

\newcommand{\Ln}{L\, \mod\, n}
\newcommand{\HS}{{\rm HS}}
\newcommand{\HSb}{\overline{\rm HS}}
\newcommand{\WZW}{{\rm WZW}}
\newcommand{\Lh}{\widehat{L}}

\newcommand{\threeb}{\overline{3}}
\newcommand{\fourb}{\overline{4}}
\newcommand{\sixb}{\overline{6}}
\newcommand{\tenb}{\overline{10}}
\newcommand{\fifteenb}{\overline{15}}
\newcommand{\nb}{\overline{n}}

\newcommand{\isi}{\stackrel{i}{=}}
\newcommand{\isone}{\stackrel{1}{=}}
\newcommand{\lessi}{\stackrel{i}{<}}
\newcommand{\lessone}{\stackrel{1}{<}}

\def\lefthook{{\vrule height5pt width0.4pt depth0pt}}
\def\righthook{{\vrule height5pt width0.4pt depth0pt}}
\def\leftrighthookfill{$\mathsurround=0pt \mathord\lefthook
     \hrulefill\mathord\righthook$}
\def\underhook#1{\vtop{\ialign{##\crcr$\hfil\displaystyle{#1}\hfil$\crcr
      \noalign{\kern-1pt\nointerlineskip\vskip2pt}
      \leftrighthookfill\crcr}}}

\newcommand{\newsection}[1]{
\vspace{12mm}
\pagebreak[3]
\addtocounter{section}{1}
\setcounter{equation}{0}
\setcounter{subsection}{0}
\setcounter{footnote}{0}
\begin{flushleft}
{\large\bf \thesection. #1}
\end{flushleft}
\nopagebreak}

\newcommand{\newsubsection}[1]{
\vspace{8mm}
\pagebreak[3]

\addtocounter{subsection}{1}
\addcontentsline{toc}{subsection}{\protect
\numberline{\arabic{section}.\arabic{subsection}}{#1}}
\noindent{\sc \thesubsection. #1}
\nopagebreak
\vspace{6mm}
\nopagebreak}

\def\ZZ{Z\!\!\!Z} 		
\newcommand{\ts}{\textstyle}

\begin{document}
\baselineskip=17pt

\hfill {ADP-96-20/M42, ITFA-96-21}

\hfill {hep-th/9607064}

\vskip 1.5cm
\begin{center}

{\Large The $\whSUn_1$ WZW Models}

\vskip 4mm

{\large Spinon Decomposition and Yangian Structure}

\vskip 1.5cm

{\large Peter Bouwknegt}

\vskip .3cm

{\sl Department of Physics and Mathematical Physics \\
     and Institute for Theoretical Physics \\ 
     University of Adelaide \\
     Adelaide, SA 5005, AUSTRALIA}

\vskip .9cm

{\large Kareljan Schoutens}

\vskip .3cm

{\sl Institute for Theoretical Physics \\
     University of Amsterdam \\
     Valckenierstraat 65 \\
     1018 XE  Amsterdam, THE NETHERLANDS}

\vskip .7cm

{\bf Abstract}

\end{center}

\baselineskip=15pt

{\small
We present a `spinon formulation' of the $\whSUn_1$
Wess-Zumino-Witten models. Central to this approach 
are a set of massless quasi-particles, called `spinons', 
which transform in the representation ${\bf \nb}$ of 
$su(n)$ and carry fractional statistics of angle 
$\theta = \pi/n$. Multi-spinon states are grouped into
irreducible representations of the yangian $Y(sl_n)$. 
We give explicit results for the $su(n)$ content of 
these yangian representations and present $N$-spinon 
cuts of the WZW character formulas. As a by-product, we
obtain closed expressions for characters of the $su(n)$ 
Haldane-Shastry spin chains.
}

\vfill

\noindent July 1996 

\newpage

\baselineskip=17pt

\newsection{Introduction}

\vskip 2mm

In this paper, we analyze the Quantum Field Theory description
of a set of mass-less `quasi-particles' in $1+1$ dimensions with the
following two defining properties
\begin{itemize}
\item
they transform in the representation ${\bf \nb}$ of an $su(n)$ symmetry 
algebra
\item
they carry fractional statistics of angle $\theta = {\pi \over n}$.
\end{itemize}
For $n=2$, such quasi-particles arise when one starts from electrons
in one spatial dimension and decouples the charge degrees of
freedom. The resulting quasi-particles, which have been called
{\em spinons}\ \cite{Ha2}, form a doublet under $su(2)$-spin and carry 
half the statistics of the original electrons, \ie, they
are {\em semions}\ with $\theta=\pi/2$. As a concrete example,
one may think of the low-lying excitations of an XXX Heisenberg 
anti-ferromagnet or of a $d=1$ Hubbard model at half filling.
We shall be using the term `$su(n)$ spinon' 
for the quasi-particles transforming in the ${\bf \nb}$
of $su(n)$ and with $\theta=\pi/n$ statistics.

The Quantum Field Theories for massless $su(n)$ spinons are 
Conformal Field Theories (CFT) with $su(n)$ symmetry. As such, they
are invariant under an affine Lie algebra $\whsun_k$, where the 
positive integer $k$ is called the level. Such theories are 
essentially unique; they are known as the $\whSUn_k$ Wess-Zumino-Witten 
(WZW) CFT. We are thus dealing with a class 
of field theories that are well-known and well-studied: detailed 
information on the spectrum and correlation functions of these 
theories is available.

The novel aspect of this work is that we shall work out a formulation 
of the $\whSUn_k$ WZW models that gives a central role to the 
fundamental quasi-particles, \ie, the $su(n)$ spinons. In particular, 
we shall interpret the (chiral) spectrum, which is usually viewed 
as a collection of irreducible Highest Weight Modules (HWM) of the 
affine symmetry, as a collection of {\em multi-spinon states}.
The motivation for such an approach comes from applications in
condensed matter physics. Let us give two examples, which both 
involve CFT's with $\whSUn_k$ symmetry.
\begin{enumerate}
\item
The {\em multi-channel Kondo effect}\ has been formulated
in terms of an $\widehat{SU(2)}_k$ CFT (this for the $s=\half$,
$k$-channel case). The essential
dynamics is the boundary scattering of the fundamental
spinons of this CFT. When $k>1$ (the multi-channel case), 
these spinons carry non-trivial topological charges, which 
are at the origin of the non Fermi-liquid behavior at low 
temperatures.
\item
In the {\em Fractional Quantum Hall effect}\, edge excitations
are described by chiral CFT's. For filling fractions $\nu$
in the so-called Jain series, $\nu = {n \over 2np + 1}$,
the edge theory is a $c=n$ CFT with extended symmetry
$\widehat{U(1)} \times \whSUn_1$. Experimentally, the
edge dynamics is probed {\em via}\ tunneling experiments.
The tunneling events can be viewed as impurity scattering 
of the fundamental quasi-particles of the CFT. For $n=1$
(and $p=1$) such a picture was used for the exact computation 
of a universal tunneling conductance curve \cite{FLS}. We 
expect that for $n>1$ a similar analysis can be done as soon as 
one can properly deal with the quasi-particles of the $\whsun_1$
symmetry, which are precisely the spinons of this paper.
\end{enumerate}

For $k>1$ the $\whSUn_k$ spinons carry additional labels
(`topological charges') related to a choice of `fusion channel' 
in multi-spinon states. We shall in this paper focus on $k=1$, where
this complication does not occur. The {\em multi-spinon}\
formalism for the $\whSUn_1$ WZW theories is very different from 
the traditional approach, which was entirely based on the affine 
symmetry $\whsun_1$.
In our new formulation, the affine symmetry is traded for an 
alternative symmetry, which is the yangian $Y(sl_n)$. In fact, the 
representation theory of the yangian quantum group can be 
interpreted as a Generalized Pauli Principle (see section 4.3) 
that guides the construction of (a basis of) multi-spinon states.

For the case $n=2$ (the original $su(2)$ spinons), a multi-spinon 
formulation has been worked out in great detail 
\cite{BPS,BLS1,BLS1a} and the connection with the representation 
theory of the yangian $Y(sl_2)$ has been made explicit.

For $n>2$, the $su(n)$ representation content of the irreducible, 
finite dimensional representations of $Y(sl_n)$ is not known
in general. We shall therefore proceed in the following indirect 
manner. We first focus on
a simpler (finite dimensional) physical system 
with exact yangian symmetry, which 
is the $su(n)$ Haldane-Shastry (HS) chain on $L$ sites.
In section 3.1 we shall give an exact and closed form result 
for the $su(n)$ content of the yangian representations that 
appear in the spectrum of these models. 
The $\whSUn_1$ WZW model arises
as the $L\rightarrow \infty$ limit of the HS model, and
our results for the yangian representations simply
carry over to the field theory (see section 3.3).

One of our main goals in this paper is to find exact
results for the $N$-spinon contribution to the CFT 
characters. Every yangian representation carries
a well-defined spinon number $N$, and it is a matter of
clever combinatorics to find the $N$-spinon cuts of
the various characters. Our strategy will be to relate
$N$-spinon characters of the $\whSUn_1$ WZW model to the
characters of a set of {\em conjugate}\ HS chains on 
$N-nl$ sites, with $l=0,1,\ldots$. Explicit results 
for the $N$-spinon cuts of various CFT characters
follow from here. They are given in section 4.4. 
It can be proved directly that the $N$-spinon characters 
sum up to the known CFT characters, which confirms 
the result of section 3.1 for the $Y(sl_n)$ characters.

We would like to stress that even for $su(2)$ the
results obtained in this paper are new. We shall find the 
following expression for the $N$-spinon contribution 
to the chiral spectrum
\be
\ch_{\WZW}^N(q,z) = q^{-{N^2 \over 4}}
 \sum_{j_1+j_2=N} \sum_{l \geq 0} (-1)^l
 {q^{\half l(l-1)} \over (q)_l}
 {q^{\half (j_1^2+j_2^2)} \over (q)_{j_1-l} (q)_{j_2-l}}
 z^{j_2-j_1} 
\ee
(see sections 4.4, 4.5 for further discussion). 

Comparing our results with quasi-particle character formulas 
proposed in \cite{FS,Ge}, we would like to stress that, where 
we work with spinons labeled by the {\em weights}\ of a 
fundamental representation of $su(n)$, \cite{FS,Ge} use 
quasi-particles labeled by the {\em roots}\ of $su(n)$. To 
illustrate the difference, let us focus on the character of the 
`vacuum module' $\Lh(\Lambda_0)$ (see section 2 for notations) 
of the $n=2$ theory. The $su(2)$ spinons act back and forth 
between the modules $\Lh(\Lambda_0)$ and $\Lh(\Lambda_1)$ and 
the character of $\Lh(\Lambda_0)$ is therefore gotten by 
restricting to even spinon-numbers
\bea
\ch_{\Lh(\Lambda_0)}(q,z) &=& 
\sum_{N \, {\rm even}} \ch_{\WZW}^N(q,z)
\nonu
&=&
 \sum_{\scriptsize{ \begin{array}{c} 
 j_1,j_2\geq 0 \\j_1+j_2 \; {\rm even} 
 \end{array}}} \, \sum_{l \geq 0} (-1)^l
 {q^{\half l(l-1)} \over (q)_l}
 {q^{\quart (j_1-j_2)^2} \over (q)_{j_1-l} (q)_{j_2-l}}
 z^{j_2-j_1} 
\nonu
&=&
 \sum_{\scriptsize{ \begin{array}{c} 
 j_1,j_2\geq 0 \\j_1+j_2 \; {\rm even} 
 \end{array}}} \, 
 {q^{\quart (j_1+j_2)^2} \over (q)_{j_1} (q)_{j_2}}
 z^{j_2-j_1} \ .
\eea
The quasi-particles of \cite{FS,Ge} act within the module 
$\Lh(\Lambda_0)$, giving a character formula with unrestricted
summations over quasi-particle numbers $r_1$ and $r_2$
\be
\ch_{\Lh(\Lambda_0)}(q,z) = \sum_{r_1,r_2\geq 0}
  {q^{r_1^2+r_2^2-r_1r_2} \over (q)_{r_1}(q)_{r_2}}
  \, z^{2r_1-2r_2} \ .
\ee
Clearly, there are many ways to interpret affine characters in 
terms of quasi-particles. From the point of view of physics, the 
choice is usually clear; while there can be many options at the 
level of kinematics, the dynamics of a given system will dictate   
a particular choice of quasi-particle basis.

In section 5 we briefly discuss other ways to set up 
a description of the $\whSUn_1$ WZW models, this time starting 
from a set of $n-1$ different quasi-particles $\phi^{(i)}$,  
$i=1,2,\ldots,n-1$, transforming in the fundamental representations 
$L(\Lambda_i)$ of $su(n)$. For example, in the case $n=3$ there will 
be two types of quasi-particles, one set transforming as the 
${\bf 3}$, the other as the ${\bf \threeb}$. In such a formulation, 
one may use a characterization of irreducible yangian representations 
in terms of parameters which are the zero's of a set of $n-1$ 
Drinfel'd polynomials (section 5.2). We shall also present examples 
of character formulas that are based on an approach that respects 
the symmetry between the conjugate representations ${\bf 3}$ and
${\bf \threeb}$ (section 5.3). 

More details, including proofs of the various statements in this
paper, will be given in a future publication \cite{BS2}. In that 
paper, we shall also present character formulas for the higher level 
($k>1$) $\whSUn_k$ WZW models, generalizing our results \cite{BLS2}, 
see also \cite{NYa,ANOT}, for the $\widehat{SU(2)}_k$ theories.

\newpage

\newsection{Basics}

\vskip -2mm

\newsubsection{$\whSUn_k$ Wess-Zumino-Witten models}

\noindent

The $\whSUn_k$ WZW models are CFT's with abundant symmetry. 
In the traditional formulation, a central role is played by 
the (bosonic) symmetry algebra $\whsun_k$, which is a so-called 
affine Lie algebra. The defining commutation relations on 
the generators $J_m^a$, $m\in \ZZ$, $a=1,\ldots, n^2-1$, are
\be
  [ J^a_m , J^b_n ] =  k \, m\, d^{ab} \delta_{m+n,0}
             + f^{ab}{}_c \, J^c_{m+n} \ .
\label{akm}
\ee
In this formula, $d^{ab}$ is a Killing metric and the $f^{ab}{}_c$ 
are the $su(n)$ structure constants, normalized as
\be
   f^a{}_{bc} f^{dbc} = -2n \, d^{ad} \ .
\ee

The Hilbert space of the $\whSUn_k$ WZW model consists
of a collection of integrable Highest Weight Modules
(HWM) $\Lh(\Lambda)$, where $\Lambda$ is an integral dominant 
weight of $su(n)$. In Dynkin notation, such weights are written as 
$\Lambda = \sum_{i=1}^{n-1} m_i \Lambda_i$, where the $\Lambda_i$ 
are the fundamental weights (we shall write $\Lambda_0=0$
for the weight of the singlet representation). For a given level 
$k$, only a finite number of HWM's are allowed. They are 
selected by the condition
\be
\sum_{i=1}^{n-1} m_i \leq k \ .
\label{cond-k}
\ee
When $k=1$, the allowed highest weights are $\Lambda_0$ and the $n-1$ 
fundamental weights of $su(n)$ (corresponding to the ${\bf 3}$, 
${\bf \threeb}$ for $su(3)$, the ${\bf 4}$, ${\bf 6}$, 
${\bf \fourb}$ for $su(4)$, etc.).

In a formulation based on the affine symmetry $\whsun_k$, the building 
blocks for the partition sum are the (chiral) characters of
$\whsun_k$, 
\be
\ch_{\Lh(\Lambda)}(q,z) = {\rm tr}_{\Lh(\Lambda)}
 ( q^{L_0} z^{\lambda} ) \ .
\ee
The notation $z^\lambda$ is shorthand for $\prod z_i^{l_i}$ where 
$\lambda = \sum l_i \Lambda_i$; this factor keeps track of the $su(n)$ 
quantum numbers of the states in the spectrum.

Related quantities are the string functions $c^\Lambda_\lambda(q)$ 
and the generating functionals $\Phi^\Lambda_\lambda(q)$.
The former record the multiplicities of $su(n)$ weights 
$\lambda$ throughout the spectrum
\be
c^\Lambda_\lambda(q) = \sum_{L_0} ({\rm mult.}\ {\rm of}\ 
 {\rm weight}\ \lambda\ {\rm at}\ {\rm energy}\ {\rm level}\ L_0) 
 \, q^{L_0}\ ,
\ee
while the latter describe the branching of the $\whsun_k$ HWM
$\Lh(\Lambda)$ into $su(n)$ irreducible representations 
(irrep) $L(\lambda)$ 
\be
\Phi^\Lambda_\lambda(q) = \sum_{L_0} ({\rm mult.}\ {\rm of}\ 
 {\rm irrep}\ L(\lambda)\ {\rm at}\ {\rm energy}\ L_0)\, 
 q^{L_0} \ .
\ee
For all these quantities, closed form expressions are known 
in the literature (see, \eg, \cite{Ka} and references
therein).

One of our goals in this paper is to recast the well-known character 
formulas for $\whsun_1$ in a way that keeps track of the 
{\em spinon-number} of the various states. This then should be viewed 
as the first step in a program where the entire analysis of the WZW 
models (characters, form factors, correlation functions) is done in a 
novel fashion: one recognizes the spinons as the fundamental elementary 
excitations and builds the theory from there. One quickly discovers 
that the affine (Kac-Moody) symmetry is not very useful for such an 
approach, as the basic multi-spinon states do not transform
nicely under the generators $J^a_m$ with $m\neq 0$. A similar
remark applies to the Virasoro symmetry. We shall therefore 
resort to an entirely different symmetry structure, which is the 
yangian $Y(sl_n)$, with generators $\{Q^a_p\}$, $p=0,1,\ldots\ $, 
together with a set of conserved charges $\{ H_q \}$, $q=1,2,\ldots\ $. 
It turns out that this symmetry is precisely tailored for a 
{\em multi-spinon formulation}\ of the $\whSUn_1$ WZW models! 
Once the level-1 case has been understood in this manner, the higher 
level models can be organized in an analogous way, see \cite{BLS2,BS2}.

The algebraic relations between the $\{Q^a_p\}$ and $\{H_q\}$ are
\be
[ Q^a_p, H_q ] = 0 \ , \qquad [ H_q, H_{q'} ] = 0 \ , 
\label{hqhh}
\ee
together with the defining relations of the yangian $Y(sl_n)$ as 
given by Drinfel'd \cite{Dr,CPa}
\bea
{\rm (Y1)} && [Q_0^a,Q_0^b] = f^{ab}{}_c \, Q_0^c \ ,
\nonu
{\rm (Y2)} && [Q_0^a,Q_1^b] = f^{ab}{}_c \, Q_1^c \ ,
\nonu
{\rm (Y3)} && [Q_1^a,[Q_1^b,Q_0^c]] 
              + ({\rm cyclic}\;\;{\rm  in}\ a,b,c)
     = A^{abc}{}_{,def} \{ Q_0^d, Q_0^e, Q_0^f \} \ ,
\nonu
{\rm (Y4)} && [[Q_1^a,Q_1^b],[Q_0^c,Q_1^d]] 
      + [[Q_1^c,Q_1^d],[Q_0^a,Q_1^b]]
\nonu
&& \qquad
     = \left( A^{ab}{}_{p,qrs} f^{cdp} + A^{cd}{}_{p,qrs} f^{abp} \right) 
       \{ Q_0^q, Q_0^r, Q_1^s \} \ ,
\nonumber 
\label{qq}
\eea
where $A^{abc,def} = {1 \over 4} f_{pqr} f^{adp} f^{beq} f^{cfr}$ 
and the curly brackets denote a completely symmetric
(weighted) product.

The yangian $Y(sl_n)$ is a non-trivial example of a quantum group, 
and as such it is equiped with a non co-commutative co-multiplication
\bea
\label{copr}
&& \Delta_{\pm}(Q_0^a) =
  Q_0^a \otimes {\bf 1} + {\bf 1} \otimes Q_0^a \ ,
\nonu
&& \Delta_\pm(Q_1^a) = 
  Q_1^a \otimes {\bf 1} + {\bf 1} \otimes Q_1^a 
  \pm \half f^a{}_{bc} Q_0^b \otimes Q_0^c \ .
\eea
The right hand sides of the relations (Y3) and (Y4) can be derived 
from the homomorphism property of this co-multiplication. For the 
case of $sl_2$, the cubic relation (Y3) is superfluous while for 
$sl_n$, with $n\geq 3$, (Y4) follows from (Y2) and (Y3).

In \cite{Sc1} it was established that the $\whSUn_1$ WZW
models carry a representation of $Y(sl_n)$, with the following 
explicit expressions for the lowest yangian generators $Q^a_0$ 
and $Q^a_1$ and charges $H_1$ and $H_2$  
\bea
&& Q_0^a = J_0^a \ , \qquad
   Q_1^a = \half \, f^a{}_{bc} \, \sum_{m>0} 
       \left( J^b_{-m} J^c_m \right) 
  \mp {n \over \strutje 2(n+2)} \, W^a_0 \ , 
\nonu
&& H_1 = L_0 \ , \qquad
   H_2 = d_{ab} \sum_{m>0} \left( m \, J^a_{-m} J^b_m \right) 
  \pm {n \over \strutje (n+1)(n+2)} \, W_0 \ .
\label{qqhh}
\eea
These expressions contain the zero modes of the following 
conformal fields (the brackets denote standard normal ordering, 
see, \eg, \cite{BS1} for conventions)
\be
W^a(z) = \half d^a{}_{bc} (J^b J^c)(z) \ , \qquad
W(z) = {1 \over 6} d_{abc} (J^a(J^bJ^c))(z) \ .
\label{WW}
\ee
The 3-index $d$-symbol that occurs in these expressions 
is completely symmetric, traceless and has been normalized 
according to
\be
  d^a{}_{bc} d^{dbc} = 
  {2(n^2-4) \over n} \, d^{ad} \ .
\ee
It was shown in \cite{Sc1} that the explicit expressions 
(\ref{qqhh}) satisfy the defining relations (\ref{hqhh}) and 
(\ref{qq}) when acting on integrable HWM's.

With this result, one immediately concludes that the HWM's
$\Lh(\Lambda)$ of $\whsun_1$ decompose into collections of 
irreducible finite-dimensional representations of $Y(sl_n)$. 
Such yangian representations have a natural interpretation in 
terms of `multi-spinon' states and each yangian representation 
carries a well-defined `spinon-number' $N$. In section 4 below we 
analyze this structure in full detail and present 
`$N$-spinon cuts' of the CFT characters.

\newsubsection{Haldane-Shastry $su(n)$ spin chains}

In 1988, Haldane \cite{Ha1} and Shastry \cite{Sh} 
proposed a class of integrable quantum spin chains that are 
different from those that can be solved by means of the Bethe 
Ansatz. A characteristic feature is that the spin-spin exchange
is not restricted to nearest neighbors. Instead, it has a
non-trivial dependence on distance, which, in the simplest
case, is of the form $1/r^2$. Here we shall recall a few 
aspects of $su(n)$ HS chains. In later sections we shall
use these results in our analysis of the $\whSUn_1$ WZW 
CFT's.

The hamiltonian $H_2$ of the $su(n)$ Haldane-Shastry (HS)
chain with $1/r^2$ exchange acts on a Hilbert space that has
$n$ states (transforming as the fundamental ${\bf \nb}$ 
of $su(n)$) for each site $i$, $i=1,2,\ldots,L$. It has the form
\be
\label{Hxxx}
H_2 = \sum_{i \neq j} \left( {z_i z_j \over z_{ij} z_{ji}}
  \right) (P_{ij}-1) \ ,
\ee
where $P_{ij}$ is a permutation operator that exchanges the
states at sites $i$ and $j$, and $z_{ij}=z_i-z_j$. We choose
the complex parameters $\{z_j\}$ as $z_j=\omega^j$, with $\omega
= \exp(2\pi i/L)$, so that the exchange described by
(\ref{Hxxx}) is proportional to the inverse-square of the chord 
distance between the sites. It was found in \cite{HHTBP} that 
the hamiltonian $H_2$ commutes with the following operators
\be
Q_0^a = \sum_i J_i^a \ , \qquad Q_1^a = 
    {\ts {1 \over 4}} \, \sum_{i\neq j}
    {(z_i+z_j) \over z_{ij}} f^{a}{}_{bc} J_i^b J_j^c \ ,
\label{q0q1}
\ee
where the $J^a_i$ are associated to the action
of $su(n)$ on the $n$ basis states at site~$i$. Furthermore,
the operators $Q^a_0$, $Q_1^a$ satisfy the defining relations
(Y1) -- (Y4) of the yangian $Y(sl_n)$. This remarkable result has 
been understood to be at the basis of the integrability of these
spin chains. Indeed, it is possible \cite{HHTBP,HT} to define mutually 
commuting integrals of motion $H_p$, $p\geq 3$, which commute 
with the hamiltonian $H_2$ and with the yangian generators. 

We thus see that the $\whSUn_1$ WZW model (which is a quantum field 
theory) and the $su(n)$ HS chain (which is a quantum mechanical 
many body system) share a common algebraic structure. This structure
was first uncovered for the HS chains \cite{HHTBP,BGHP}, and later 
established for the CFT counterparts \cite{BPS,BLS1}.

\vskip 4mm

\noindent
We shall now briefly summarize the solution of the HS chain 
as given in \cite{Ha2,HHTBP,BGHP}. In this work, eigenstates  of the 
hamiltonian (\ref{Hxxx}) are characterized {\em via}\ `motifs',
which are sequences of $L+1$ digits `0' or `1', beginning and ending 
with `0', and containing at the most $n-1$ consecutive `1'. The
motifs can be parametrized by `rapidity' sequences 
$l_j \in \{1,2,\ldots,L-1 \}$ that indicate the positions of 
the `1'. Every motif corresponds to a degenerate set of 
eigenstates, which together form an irreducible representation 
of the yangian symmetry. These states have $H_2$ eigenvalue
\be
  H_2 = - \sum_j l_j (L-l_j) 
\ee
and crystal momentum $K= - \left( {2\pi \over L} H_1 \right) 
\mod \, 2\pi$, where
\be
  H_1 = - \sum_j l_j \ .
\ee

The CFT analogue of the operator $H_1$ will be the Virasoro 
zero mode $L_0$, which is interpreted as `energy' in the CFT 
setting. One easily finds that the lowest possible
value of $H_1$ on the $L$-site chain is 
\be
  H_1^{(0)} =  - \left( {n-1 \over 2n} \right) L^2
           + \half | \Lambda_k |^2 \ ,
\label{H1GS}
\ee
where $L \equiv k \; \mod \, n$ with $k=0,1,\ldots,n-1$
and $|\Lambda_k|^2 = {k(n-k) \over n}$. 

For $n>2$, the $su(n)$ content of yangian irreducible representations 
is not known in general. The $su(n)$ content of the motif-related
yangian irreducible representations was first studied \cite{HHTBP,HH}, 
where the following procedure was proposed. One replaces every
`0' by `)(' (a `(' for the first `0' and `)' for the last), so that the 
motif breaks up as a string of elementary motifs of the form 
$(11\ldots 1)$ with at most $n-1$ `1's. These elementary motifs 
correspond to the singlet ($n-1$ `1's) and to the $n-1$ fundamental 
irreducible representations of $su(n)$. For example, for $su(3)$ one has 
that `(11)' is a singlet, `(1)' is a ${\bf \threeb}$ and `()' is a 
${\bf 3}$. 
The tensor product of the elementary motifs gives an upper bound to the 
total $su(n)$ content of the motif. The precise content is obtained 
from here by taking into account certain reductions. For $su(2)$ these 
reduction amount to the total symmetrization of adjacent doublets,
\ie, `()()' represents a triplet rather than the product of
two doublets. The reductions that are needed for $su(3)$ have
been exemplified in \cite{HH}, but until now have not been given 
in closed form. In \cite{HH}, an indirect characterization 
of the motif-related yangian representations was given using 
`squeezed strings'. This was used to show (for $n=2$ and
numerically for $n$ up to 6 and small $L$) that the motif 
prescription is complete in the sense that it accounts for the 
correct number of $n^L$ eigenstates of the HS hamiltonian. 

In section 3.1 below we shall give a very simple and general 
characterization of the $su(n)$ content of the motif-related 
yangian irreducible representations, and from there derive a simple 
formula for the HS character $\ch_{\HS}^L(q,z)$, defined as
\be          
\ch_{\HS}^L(q,z) \equiv 
\, {\rm tr}_{\HS} \left( q^{L_0} 
\, z^\lambda \right) \ ,
\label{ZHS}
\ee
where (see (\ref{H1GS}))
\be
L_0 = H_1 - H_1^{(0)} + {1\over2} |\Lambda_k|^2 \ .
\label{ZHSa}
\ee

\newpage

\newsection{$\whSUn_1$ CFT as a limit of the $su(n)$ HS chain}

\vskip -2mm

\newsubsection{Yangian irreducible representations, I}

In this subsection, we give a closed form result for the $su(n)$ 
content of yangian irreducible representation that
corresponds to a general motif of the $L$-site HS
chain. We encode the motif by an ordered set
$1 \leq l_1 < l_2 < \ldots < l_s \leq L-1$, the
$l_j$ denoting the positions of the `1' in the motif.
We shall express the yangian character in the 
characters of $su(n)$ irreducible representations with Dynkin 
labels $(m \, 0 \ldots \, 0)$, which we write as $\chi_m(z)$.

By working out some examples, one quickly discovers
that the motif `00 \ldots 0' ($L+1$ zeros) represents a 
completely symmetrized tensor product $({\bf n}^{\otimes L})_s$
with character $\chi_L(z)$. 
Replacing one of the `0' by `1', \ie,
introducing a single non-zero $l$, represents
the product $\chi_{L-l}(z) \chi_l(z)$ {\em minus} the
completely symmetric term $\chi_L(z)$. 

This pattern can quickly be generalized to a general motif 
with `1' at positions $l_1<l_2< \ldots <l_s$, leading to an 
alternating sum over ordered subsets $S'=\{ l_{i_1},l_{i_2},
\ldots,l_{i_t} \}$ of $S=\{l_1,l_2,\ldots,l_s\}$. 
We therefore propose the following closed expression for the 
character $\ch^Y_{\{l_j\}}(z)$ of the
motif-associated yangian irreducible representations
\be
\chi^Y_{ \{l_j\} }(z) = 
  \sum_{S' \subset S} (-1)^{s-t}
    \chi_{L-l_{i_t}}(z) \chi_{l_{i_t}-l_{i_{t-1}}}(z)
    \ldots \chi_{l_{i_1}}(z) 
\label{yangchar}
\ee
(cf.\ \cite{CPb}, Proposition 4.6, for $Y(sl_2)$).

Notice that this expression has the effect of anti-symmetrizing 
the tensor power ${\bf \nb}^{\otimes (p+1)}$ of adjacent $n$-plets
separated by $p$ `1's. For $p=n-1$ this leads to a singlet since
\be
  \sum_{S' \subset \{1,2,\ldots,n-1 \} } (-1)^{n-1-t}
    \chi_{n-l_t}(z) \chi_{l_t-l_{t-1}}(z)
    \ldots \chi_{l_1}(z) = \chi_0(z) =1 \ ,
\ee
while for $p \geq n$ consecutive `1' the result is identically
zero. This then confirms the observation of \cite{HHTBP} that
non-vanishing motifs can have no more than $n-1$ consecutive `1'.

\newsubsection{HS characters}

A direct consequence of the result (\ref{yangchar}) is 
that the sum over all motifs of the associated yangian
characters simply gives $(\chi_1)^L$, \ie, the $su(n)$
character of the tensor product of $L$ spins in the ${\bf n}$
of $su(n)$. We can easily do better than this, and
evaluate the sum over all motifs keeping track of the 
eigenvalues $H_1$, meaning that we are evaluating 
the character (\ref{ZHS}). Consider $su(3)$, $L=3$ as 
an example. We have

\vskip 4mm
\begin{center}
\begin{tabular}{cccc}
motif & $ \quad q^{H_1+3} \quad $ & character & $su(3)$ content
\stru \\ \hline
$0110$ & $ q^0$ & $ 
        \chi_1^3 - \chi_2\chi_1 - \chi_1\chi_2 + \chi_3 $ & {\bf 1}
\stru \\
$0010$ & $ q^1$ & $ \chi_1 \chi_2 - \chi_3  $ & {\bf 8}
\stru \\
$0100$ & $ q^2$ & $ \chi_2 \chi_1 - \chi_3 $ & {\bf 8}
\stru \\
$0000$ & $ q^3   $ & $ \chi_3 $ & {\bf 10}
\stru
\end{tabular}
\end{center}
\vskip 4mm

The $q$-powers can be written as $q^{\sum l'_j}$ where $l'_j$ 
are the positions of the `0' ($\neq 0,L$) in the motif. We can 
write them as $\prod_j \left(1-(1-q^{l'_j})\right)$ and then 
perform the sum over all motifs. In the example this gives
\be
 \left( (1-q)(1-q^2) \, \chi_3 
- (1-q^2)  \, \chi_2 \chi_1 
- (1-q) \, \chi_1 \chi_2 + \chi_1^3 \right) \ . 
\ee
In the general case, one finds that the total coefficient of 
\be
\chi_{L-l_s}(z) \chi_{l_s-l_{s-1}}(z) \ldots \chi_{l_1}(z)
\ee
in the character equals
\be
q^{{n-1 \over 2n}L^2 -\half L(L-1)} (-1)^{L-1-s} 
\prod_{i=1}^s (1-q^{l'_i}) \ ,  
\ee
where the set $\{l'_j\}$ is the complement of $\{l_j\}$
in $\{1,2,\ldots,L-1\}$. This gives
\bea
\lefteqn{\ch_{\HS}^L(q,z) =}
\nonu
&&  q^{{n-1 \over 2n}L^2 - \half L(L-1)} 
\sum_{1 \leq l_1 < \ldots < l_s \leq L-1}
(-1)^{L-1-s} \, { (q)_{L-1} \over \prod_{i=1}^s (1-q^{l_i}) } 
\chi_{L-l_s}(z) \chi_{l_s-l_{s-1}}(z) \ldots \chi_{l_1}(z) \ ,
\nonu
&&
\eea
where
\be
(q)_M \equiv \prod_{l=1}^M \ (1-q^l) \ .
\label{qnum}
\ee

Expanding this result in powers of $q^{-1}$ leads to
a representation of the character in terms of
occupation numbers $m_k=0,1,\ldots $ of `orbitals' of 
`energy' $k=0,1,2,\ldots$
\be
\ch_{\HS}^L(q,z) = (-1)^L \, q^{{n-1 \over 2n}L^2 -\half L(L+1)} \, (q)_L
 \sum_{\sum_k m_k = L}
 q^{-\sum_k k m_k} \prod \chi_{m_k}(z) \ .
\label{qinverse}
\ee

One would like to view the HS character as a $q$-deformation 
of a formula that gives the decomposition of the $su(n)$ character 
of the full Hilbert space in terms of irreducible representations 
of $su(n)$. To write such a formula, we need to introduce a
bit of notation. We denote by $\eps^i$, $i=1,2,\ldots,n$ 
the weights of the fundamental representation $L(\Lambda_1)$ 
(the ${\bf n}$) of $su(n)$, with inner product $(\eps^i,\eps^j) =
\delta^{ij}-{1 \over n}$. By $P$ we denote the weight space
of $su(n)$ and $P^{(k)}$ denotes the weights in the 
conjugacy class $k$, \ie, if $\lambda = \sum_i m_i \Lambda_i$, then 
$\lambda \in P^{(k)}$ iff $\sum im_i \equiv k \, \mod \, n$. 
By multinomial expansion we have
\be
\left( \chi_1(z) \right)^L
= \sum_{\lambda \in P^{(\Ln)}} z^\lambda \, 
  M_{\lambda}^{L,n}(q=1) \ ,
\label{decomp}
\ee
where, for $\lambda \in P^{(\Ln)}$,
\be
M_{\lambda}^{L,n}(q) \equiv 
\left[ \begin{array}{c} L \\
  {L \over n} + (\lambda,\eps^1), \ldots ,  
  {L \over n} + (\lambda,\eps^n) \end{array} 
\right]_q
\ee
with the $q$-multinomial defined as
\be
\left[ \begin{array}{c} m_1+m_2+ \ldots + m_n \\
  m_1 , m_2 , \ldots , m_n \end{array} \right]_q
  = { (q)_{m_1+m_2+\ldots+m_n} \over
    (q)_{m_1} (q)_{m_2} \ldots (q)_{m_n} } \ .
\ee

As announced, we can write the HS character
as a natural $q$-deformation of (\ref{decomp}).
{}From (\ref{qinverse}) we derive 
\be
\ch_{\HS}^L(q,z) = 
  \sum_{\lambda \in P^{(\Ln)}} 
  z^\lambda \, q^{{n-1 \over 2n}L^2} \,
  M_{\lambda}^{L,n}(q^{-1})  
\ee
and it follows that
\be
\ch_{\HS}^L(q,z) = \sum_{\lambda \in P^{(\Ln)}} 
  z^\lambda \,
  q^{\half |\lambda|^2} \, 
  M_{\lambda}^{L,n}(q) \ .  
\label{HSchar}
\ee
Note that the correct ground state energy (\ref{H1GS}) 
follows directly from this character. For $n=2$, a formula 
of this type was already given in \cite{Me}.

\newsubsection{The limit $L\to\infty$: CFT characters}

Having obtained explicit results for the HS characters,
we can take the limit \hbox{$L\rightarrow \infty$}. In
this limit, the combination $L_0$ in (\ref{ZHSa})
corresponds to the Virasoro zero mode $L_0$ of the CFT, and 
the HS characters directly turn into CFT characters.
For on the string function $c_\lambda^\Lambda(q)$ 
($\Lambda=\Lambda_{\Ln}$), we find
\be
c_\lambda^\Lambda(q) 
= \lim_{L\rightarrow \infty}
  q^{\half |\lambda|^2} \, M_{\lambda}^{L,n}(q) 
= {q^{\half |\lambda|^2} \over 
    (q)_\infty^{n-1} } \ ,
\ee
in agreement with the known result.

Closely related to the string functions are the generating 
functionals $\Phi_{\HS,\lambda}^L(q)$ defined as
\be
\Phi_{\HS,\lambda}^L(q) = 
\sum_{L_0} ({\rm mult.}\ {\rm of}\ {\rm irrep}\ L(\lambda)\
 {\rm at}\ L_0) \, q^{L_0}
\ee
(with $L_0$ as in (\ref{ZHSa})) for which we derive from 
(\ref{HSchar}) (see \cite{BS2} for details)
\bea
\Phi_{\HS,\lambda}^L(q) &=& 
{ (q)_L \over
\prod_{i=1}^n (q)_{ {L \over n} + (\lambda,\eps^i) + (n-i)} } \,
q^{\half | \lambda |^2}
\prod_{\alpha \in \Delta_+}
\left( 1- q^{(\lambda+\rho,\alpha)} \right) 
\nonu
&=& q^{ - {L(L-n)\over 2n}} \, 
K_{( {L\over n} + (\lambda,\epsilon^1), \ldots,
  {L\over n} + (\lambda,\epsilon^n)),(1^L) }(q) \ ,
\label{PhiHS}
\eea
where $K_{\lambda,\mu}(q)$ is the so-called Kostka polynomial 
(see, \eg, \cite{Ki} and references therein). In this expression 
$\Delta_+$ stands for the positive root lattice and 
$\rho = \half \sum_{\Delta_+} \alpha$.

In the limit $L \rightarrow \infty$, the $\Phi_{\HS,\lambda}^L(q)$ 
go to the generating functionals $\Phi^{\Lambda}_{\lambda}(q)$
of the $\whSUn_1$ WZW CFT, where $\Lambda = \Lambda_{\Ln}$. 
For $n=2$, $\Phi^{\Lambda}_{\lambda}(q)$ are irreducible
characters of a $c=1$ Virasoro algebra, while for $n>2$ the
$\Phi^{\Lambda}_{\lambda}(q)$ are irreducible characters
of the $\cW_n$ algebra at $c=n-1$. Indeed, in the limit 
$L \rightarrow \infty$ (\ref{PhiHS}) reduce to
\be
\lim_{L\rightarrow\infty} \Phi_{\HS,\lambda}^L = 
{ 1 \over (q)_\infty^{n-1} } \,
q^{\half | \lambda |^2}
\prod_{\alpha \in \Delta_+}
\left( 1- q^{(\lambda+\rho,\alpha)} \right) \ ,
\label{PhiCFT}
\ee
in agreement with the known character formulas for
the $\cW_n$ algebra at $c=n-1$ \cite{BS1}.

Note that eqn. (\ref{PhiCFT}) is a special case of the statement 
that the $\widehat{sl_n}/sl_n$ characters (at arbitrary level) can 
be written as limits of $q^{-{kL(L-n)\over 2n}}$ times Kostka 
polynomials. This result was conjectured in \cite{Ki} and recently 
proven in \cite{NY}.

\newpage

\newsection{$N$-spinon states in the $\whSUn_1$ CFT}

\vskip -2mm

\newsubsection{Spinons}

In the previous section, we have seen that 
the $\whSUn_1$ WZW CFT can be viewed as the limit $L\rightarrow \infty$
of the HS quantum chain. Under this correspondence, the CFT primary
field transforming as the ${\bf \nb}$ of $su(n)$ corresponds to
the elementary `spinon' excitation over the HS ground state.

Let us explain the nature of these spinons. For the purpose
of illustration, we put $n=3$ and consider the $su(3)$ HS chain on 
$L$ sites, where we assume that $L$ is a multiple of 3. The motif 
\be
     0110110\ldots 110 \rightarrow (11)(11)\ldots (11)
\ee
represents the ground state, which is a singlet under $su(3)$. The
idea is that the fundamental spins (each carrying a ${\bf 3}$ of 
$su(3)$) have bound into groups of three that each carry the 
singlet representation, which is the fully anti-symmetrized
product of the constituent ${\bf 3}$. The lowest excited states are 
obtained by removing a single `1' from the motif and, possibly, 
shifting one or more of the remaining `1'. This typically leads
to a motif containing three times `(1)' and for the rest `(11)'.
The elementary motif `(1)' represents the anti-symmetrized product
of two ${\bf 3}$, which is a ${\bf \threeb}$. The fact that removing a 
single `1' from a motif leads to {\em three}\ independent excitations 
`(1)' can be seen as a manifestation of the fact that the elementary
excitation `(1)' carries `fractional statistics' with parameter
$\theta = \pi/3$. The generic 3-spinon excitation is of the form
(the dots stand for groups `(11)')
\be
\ldots (1) \ldots (1) \ldots (1) \ldots 
\ee
and has $su(3)$ content ${\bf \threeb}^{\otimes 3} = 
{\bf 1} \oplus {\bf 8}_2 \oplus {\bf \tenb}$. 
Two `(1)' motifs may merge into `()' giving
\be
\ldots (1) \ldots () \ldots \ , \qquad {\rm or}\ \qquad
\ldots () \ldots (1) \ldots 
\ee
with $su(3)$ content ${\bf 3} \otimes {\bf \threeb} = 
{\bf 1} \oplus {\bf 8}$ for the case where the `(1)' and `()' 
are separated by at least one singlet group `(11)'. When they 
are adjacent there is a further reduction to
\be
\ldots (10) \ldots \ , \qquad {\rm or}\ \qquad
\ldots (01) \ldots 
\ee
with $su(3)$ content ${\bf 3} \otimes {\bf \threeb} - {\bf 1} 
= {\bf 8}$. Note that we have used the notation `(10)' for 
`(1)()' and `(01)' for `()(1)' (compare with appendix~C).

The fact that the total number of $3^3=27$ possible 3-spinon 
states gets reduced when the 3 individual spinons occupy nearby 
orbitals follows from a Generalized Pauli Principle, see section 
4.3. For $n=2$ there is only one non-trivial elementary motif, 
the `()' representing an $su(2)$ doublet,
and the Generalized Pauli Principle amounts to symmetrizing
the product ${\bf 2}^{\otimes l}$ of $l$ adjacent `()'.

For general $n$, the elementary excitations
`(11 \ldots 1)' ($n-2$ times `1'), which we shall call 
{\em spinons}, have the two characteristic features
which we already mentioned in our introduction: they 
transform in the $su(n)$ representation ${\bf \nb}$
and carry fractional statistics of angle $\theta = {\pi \over n}$.
A generic $N$-spinon state will carry the $su(n)$ representation
${\bf \nb}^{\otimes N}$ but in many cases the actual $su(n)$ content 
is reduced according to a Generalized Pauli Principle.
The $n=3$ example shows that the correct rules for general $n$ 
are rather subtle. A detailed description shall be given in 
section 4.3 below.

\newsubsection{Multi-spinon states}

In the $\whSUn_1$ WZW CFT the spinon excitations are
represented by the primary field $\phi^\alpha(z)$,
$\alpha = 1,2,\ldots,n$, transforming 
in the ${\bf \nb}$ of $su(n)$ and having conformal dimension 
$h = {n-1 \over 2n}$. The CFT ground state corresponds to
a half-infinite sequence of vacuum motifs 
$(11\ldots1)(11\ldots1)\ldots$\ . Single-spinon excitations
are created by applying a positive-energy mode of the spinon 
field $\phi^\alpha(z)$ to the vacuum. Such modes are
defined by
\be
\phi^\alpha(z) = \sum_{m \in \ZZ} \phi^\alpha_{-{n-2k-1 \over 2n}-m} 
z^{m-{k \over n}} \ ,
\ee
for modes acting on the HWM (`sector') with highest weight
$\Lambda_0$ for $k=0$ and $\Lambda_{n-k}$ for $k=1,\ldots,n-1$.

The idea is now that the entire (chiral) Hilbert space
of the CFT is spanned by {\em multi-spinon} excitations,
which are created by repeatedly acting with modes of the
fundamental spinon fields. A general $N$-spinon state is
written as
\bea
&&    \phi^{\alpha_N}_{-{n-(2N-1) \over 2n}-n_N}  \ldots 
      \phi^{\alpha_3}_{-{n-5 \over 2n}     -n_3} 
      \phi^{\alpha_2}_{-{n-3 \over 2n}     -n_2}
      \phi^{\alpha_1}_{-{n-1 \over 2n}     -n_1} |0 \rangle \, ,
\nonumber \\[4mm]
&& \qquad \qquad \qquad \qquad 
   {\rm with} \;\; n_N \geq n_{N-1} \geq \ldots
    \geq n_2 \geq n_1 \geq 0 \ .
\label{Nsp}
\eea
The fact that states with different ordering for the $n_i$ 
can be expressed in terms of states (\ref{Nsp}) follows from
the Generalized Commutation Relations (GCR) of the spinon modes,
see below. The energy (eigenvalue of the Virasoro zero mode 
$L_0$) of these states is 
\be
   L_0 = -{ N(N-n) \over 2n } + \sum_{i=1}^N n_i \ .
\label{Lzero}
\ee

The yangian generators $Q_i^a$ and the conserved charge $H_2$ 
have a simple action on one-spinon states. Having assigned the
$su(n)$ representation ${\bf \nb}$ to the fundamental spinons,
we should take the lower sign in (\ref{qqhh}). (This choice
is such that the space of $N$-spinon states will be invariant
under the action of the yangian.) We then find 
\bea
&&
Q_0^a \, \phi^\alpha_{-{n-1 \over 2n}-n_1} \vac
= (t^a)^\alpha{}_\beta \, \phi^\beta_{-{n-1 \over 2n}-n_1} \vac
\ , 
\nonu
&&
Q_1^a \, \phi^\alpha_{-{n-1 \over 2n}-n_1} \vac
= \left[ - n \, n_1- {n-2 \over 4} \right] (t^a)^\alpha{}_\beta \, 
\phi^\beta_{-{n-1 \over 2n}-n_1} \vac
\label{q1}
\eea
and
\be
H_2 \, \phi^\alpha_{-{n-1 \over 2n}-n_1} \vac
= E_2(n_1) \,
  \phi^\alpha_{-{n-1 \over 2n}-n_1} \vac \ ,
\label{h2}
\ee
where 
\be
E_2(n_1) = \left[ {(n-1)(n-2) \over 6n}
         + n \, n_1^2 + (n-1) \, n_1 \right] .
\ee
In these formulas, the matrices $(t^a)^\alpha{}_\beta$
describe the representation ${\bf \nb}$ of $su(n)$ and 
we have
\be
t^a t^b = {1 \over n} d^{ab} - \half f^{ab}{}_c t^c
         - \half d^{ab}{}_c t^c \ .
\ee
In technical terms, the one-spinon states constitute an
{\em evaluation representation}\ of the yangian $Y(sl_n)$.

The action (\ref{q1}) of the yangian can be extended 
without problem to the level of multi-spinon states 
of the form (\ref{Nsp}). One then finds that the yangian 
generators preserve the spinon-number, \ie, they map an 
$N$-spinon state of the form (\ref{Nsp}), with modes 
$\{ n_i \}$, $i=1,2,\ldots,N$, onto similar states with
modes $\{ n'_i \}$ that are equal to or smaller than
(in the sense of a natural partial ordering $\preceq$) 
the original modes $\{ n_i \}$. A similar result is
found for the action of $H_2$. As a direct consequence,
one finds that one may define linear combinations
of states of the form (\ref{Nsp}), with labels
$\{ n'_i \} \preceq\{ n_i \}$,  such that on the new
states $H_2$ is diagonal with eigenvalue
\be
  H_2 = \sum_{i=1}^N E_2(n_i-{i-1 \over n}) \ .
\label{htwo}
\ee
The eigenspaces of $H_2$ naturally carry a representation
of the yangian, and in this manner we arrive at a
decomposition of the space of $N$-spinon states in terms 
of irreducible representations of the yangian $Y(sl_n)$. 
For $n=2$ this procedure has been worked out in great detail in 
\cite{BPS} and in our papers \cite{BLS1,BLS1a} with 
A.W.W.~Ludwig. See also \cite{AN} where a number of results 
for general $n$ were given.

Spinon fields are not free fields, and we already remarked
that they carry statistics which are neither bosonic nor 
fermionic. To illustrate this point we remark that
the Operator Product Expansion of spinon fields 
$\phi(z)$ and $\phi(w)$ has the form
\be
\phi(z) \phi(w) = (z-w)^{-{1 \over n}} 
  \left[ \phi'(w) + \ldots \right] ,
\ee
where $\phi'(z)$ is the primary field in the representation 
$L(\Lambda_{n-2})$ of $su(n)$ (for $n=2$ this is the identity).
It is the value $-{1 \over n}$ of the exponent which leads to 
statistics of angle $\pi/n$. Note that it is {\em not}\ the 
conformal dimension $h={n-1 \over 2n}$ which determines the 
statistics of the spinons. Indeed, while for $n\rightarrow\infty$ 
this dimension approaches the value $h=\half$ of a free fermion, 
the statistics approach those of a bosonic theory.

Following \cite{Ha3}, we may interpret the non-trivial $\pi/n$
statistics of the spinons as `exclusion statistics'.
As such, they manifest themselves in the motif description
of the spectrum (`creating a single hole leads to an $n$-spinon 
excitation', see section 4.1) or, more directly, in the fact that 
the minimal allowed energy difference between adjacent spinon modes 
is ${1 \over n}$ (compare with (\ref{Nsp})).

Due to the statistics aspect, it is non-trivial to compare 
different multi-spinon states and to select a set that forms 
a basis of the chiral Hilbert-space (the $\cW_n$ modules, 
say) of our CFT. A direct but cumbersome way to work out relations 
among multi-spinon states is by using `Generalized Commutation
Relations' (GCR) for the modes of the fundamental spinon
fields. In our papers \cite{BLS1,BLS1a} with A.W.W.~Ludwig 
we derived the fundamental GCR for the case $n=2$, $k=1$ and 
worked out the ensuing relations among multi-spinon states.  

In this paper we shall proceed at a more abstract level and 
proceed by giving a relation between the $N$-spinon states 
of the $\whSUn_1$ WZW model and associated $N-nl$-site 
($l=0,1,\ldots$) HS models with fundamental spins in the 
${\bf \nb}$ of $su(n)$. In this way, we shall be able to 
characterize a basis of multi-spinon states and compute 
$N$-spinon cuts of a variety of CFT characters.

\newsubsection{Yangian irreducible representations, II}

The $su(n)$ content for a multi-spinon yangian multiplet 
is quickly found if we remember that the CFT can be viewed as 
the $L\rightarrow \infty$ limit of an $L$-site HS chain. Using 
the following rule we can construct a motif of the $L=\infty$ 
HS chain from the modes sequence $n_1 \leq n_2 \leq \ldots \leq n_N$ 
of a multi-spinon state. 

We restrict to sequences $n_i$ with at most $n-1$ consecutive 
$n_i$ equal (the other ones are either zero or equal to a state 
of lower spinon number). Consider the integers $k=0,1,\ldots$ 
and denote by $m_k$ the number of $n_i$ equal to $k$. Next, 
replace $m_k$ by the elementary motif $(11\ldots1)$ having
$n-1-k$ times `1', and replace `)(' and the initial `('by 
`0'. In this way, one finds a half-infinite motif of 
the $L=\infty$ chain, which differs from the vacuum motif 
(in the $\Lh(\Lambda_{-N\, {\rm mod}\, n})$ sector)
in finitely many places. The $su(n)$ content of the multi-spinon 
yangian multiplet is then given by (\ref{yangchar}), restricted
to the (finite) non-trivial part of the motif.

As an example, we again put $n=3$ and consider the 3-spinon states 
with modes $n_1,n_2,n_3$. One quickly finds the result shown
in Table 1%
\footnote{We use the notation $a \isi b$ for $b-a=i$ and
          $a \lessi b$ for $b-a > i$.}
(compare with section 4.1).

\begin{table}
\begin{center}
\begin{tabular}{ccc}
\stru modes & motif & $su(3)$ content  \\ \hline
\stru $n_1 < n_2 < n_3 \;$  & $\ldots (1) \ldots (1) \ldots (1) \ldots$  & 
                      ${\bf 1} \oplus {\bf 8}_2 \oplus {\bf \tenb}$
\\
\stru $n_1 = n_2 \lessone n_3 \;$ & $\ldots () \ldots (1) \ldots$ & 
                      ${\bf 1} \oplus {\bf 8}$
\\
\stru $n_1 \lessone n_2 = n_3 \;$ & $\ldots (1) \ldots () \ldots$ & 
                      ${\bf 1} \oplus {\bf 8}$
\\
\stru $n_1 = n_2 \isone n_3 \;$ & $\ldots (01) \ldots$ & {\bf 8}
\\
\stru $n_1 \isone n_2 = n_3 \;$ & $\ldots (10) \ldots$ & {\bf 8}
\end{tabular}
\end{center}
\caption{3-Spinon states in the $\widehat{SU(3)}_1$ theory.}
\end{table}

Before coming to $N$-spinon characters, we wish to rephrase the 
$su(n)$ structure of the yangian multiplet with mode sequence 
$n_1 \leq n_2 \leq \ldots \leq n_N$ in a way that is more convenient 
for the CFT setting. 
The intuition for this rephrasing comes from the 
structure of the GCR for spinon modes $\phi^\alpha_r$, which 
roughly speaking tell us two things. First of all, they imply 
that we should antisymmetrize over neighboring spinons with equal 
mode-indices, and secondly they imply that in certain cases the 
singlet combination (`trace') of $n$ spinons has to be subtracted.

We start from a mode sequence $0\leq n_1 \leq n_2 \leq
\ldots \leq n_N$. We denote by 
\be
\chih_{\{n_i\}}(z)
\label{chih}
\ee
the character of ${\bf \nb}^{\otimes N}$, antisymmetrized over equal 
$n_i=n_{i+1}=\ldots = n_{i+k}$. Next, we have a prescription 
(see Appendix~B) for reducing this character further by placing 
$l$ `hooks', which each eliminate $n$ modes $n_i$ from the 
sequence. In the character, placing a hook means that the 
corresponding $n$ spinons have combined into a singlet and no 
longer contribute. We write the reduced character obtained by 
placing $l$ hooks in all possible positions as
\be
\chih^{(l)}_{\{n_i\}}(z) \ .
\ee
We claim the following result for the $N$-spinon yangian 
character $\chih^Y_{\{n_i\}}(z)$
\be
\chih^Y_{\{n_i\}}(z) = 
  \sum_{l} (-1)^{l} \chih^{(l)}_{\{n_i\}}(z) \ .
\label{yangchar2}
\ee
Clearly, the two steps in this little construction reflect
the two aspects (`antisymmetrization' and `trace subtraction') of 
the GCR. 

One may view the rule encompassed in the formula (\ref{yangchar2}) 
as a {\em Generalized Pauli Principle}\ for the construction of
multi-spinon states in the $\whSUn_1$ CFT. Returning 
to the example $su(3)$, $N=3$ (3 spinons), we can see this 
Generalized Pauli Principle in action. Focusing on 
$(n_1,n_2,n_3) = (m,m,m+1)$, corresponding to the motif 
`$\ldots(01)\ldots$', we have
\bea
&& \chih_{m,m,m+1}(z) = (\chi_1\chi_1-\chi_2)(z) \, \chi_1(z) 
\nonumber \\[3mm]
&& \chih^{(0)}_{m,m,m+1}(z) = (\chi_1^3 - \chi_2\chi_1)(z)\ ,
\quad
\chih^{(1)}_{m,m,m+1}(z) = 
\chih_{\scriptsize{\underhook{m,m,m+1}}}(z) = \chi_0(z)
\nonu
&& \chih^Y_{m,m,m+1}(z) = (\chi_1^3 - \chi_2\chi_1 - \chi_0)(z) \ ,
\label{char01}
\eea
confirming that the $su(3)$ content of this yangian 
multiplet is a single ${\bf 8}$.

One may note that the leading term (with hook-number $l=0$) 
of (\ref{yangchar2}) overlaps with states at lower spinon
number. This redundancy is corrected for by subtractions
at higher hook-numbers. One quickly finds that the lowest 
$N$-spinon state in the CFT spectrum, for $N$ of the form 
$N=p(n-1)+q$ with $q\in\{0,1,\ldots,n-2\}$, has Dynkin labels 
$m_1=p$ and (for $q\neq 0$) $m_{n-q}=1$ and energy
\be
L_0^{({\rm min},N)} 
   = {1\over 2} | p\Lambda_1 + \Lambda_{n-q}|^2 
   = {N^2 \over 2n(n-1)} + {q(n-1-q) \over 2(n-1)} \ .
\label{L0min}
\ee
In Appendix~A we illustrate the spinon decomposition of the WZW 
spectrum for the case $\widehat{SU(3)}_1$.
 
\newsubsection{$N$-spinon characters}

Having come this far, we are finally ready to evaluate
the $N$-spinon contributions to the various characters of
the $\whSUn_1$ CFT. We write the $\whsun_1$ characters
as
\be
  \ch_{\Lh(\Lambda_{n-k})}(q,z) = \sum_{N \equiv k \,\mod\, n}         
                              \ch_{\WZW}^N(q,z) \ ,
\ee
where $\ch_{\WZW}^N(q,z)$ denotes the total contribution of the 
$N$-spinon states to the chiral spectrum of the CFT. It can be 
written as
\be
\ch_{\WZW}^N(q,z) = 
  \sum_{0\leq n_1\leq n_2\leq \ldots\leq n_N}
  q^{-{N(N-n) \over 2n} + \sum_i n_i} 
  \, \chih^Y_{n_1,\ldots,n_N}(z) \ .
\label{CFTchar}
\ee
Using (\ref{yangchar2}), we can write this summation as
an alternating sum over a hook-number $l$. If we then evaluate
the character at fixed $l$, we find the following beautiful
result
\bea
\ch_{\WZW}^{N,l}(q,z) 
&=& \sum_{n_1\leq n_2\leq \ldots\leq n_N}
 q^{-{N(N-n) \over 2n} + \sum_i n_i} 
 \chih^{(l)}_{n_1,\ldots,n_N}(z) 
\nonu
&=& {q^{\half l(l-1)} \over (q)_{l}} \, 
    {1 \over (q)_{N-n l}} \ch_{\HSb}^{N-n l}(q,z) \ ,
\label{ZNl}
\eea
so that
\be
\ch_{\WZW}^N(q,z) = \sum_{l\geq 0} (-1)^l
{q^{\half l(l-1)} \over (q)_{l}} 
{1 \over (q)_{N-n l}} \ch_{\HSb}^{N-n l}(q,z) \ .
\label{WZWN}
\ee
Here $\ch_{\HSb}^{N-nl}(q,z)$ denotes the character of a
{\em conjugate}\ HS model, with fundamental spins in the
${\bf \nb}$ of $su(n)$, on $L=N-nl$ sites.

The result (\ref{ZNl}) can be explained as follows. Group 
theoretically, an $N$-spinon state with $l$ hooks corresponds 
to a combination of $N-n l$ spinons that each transform in
the ${\bf \nb}$ of $su(n)$. This is the same kinematics as that 
of the conjugate HS model on $N-n l$ sites, and it can be 
shown that the systematics of how the fundamental spins 
combine is identical in both cases. 
The identity (\ref{ZNl}) hinges on the following relation 
between the yangian character $\overline{\chi}^Y_{\{b_i\}}(z)$
for a motif `$0 \, b_{N-1} \ldots \, b_2 b_1 0$' 
(with $b_i=0$ or $1$) for the conjugate HS chain on $N$ 
sites and the anti-symmetrized $N$-spinon character 
$\chih_{\{n_i\}}(z)$ (\ref{chih}). Using (\ref{yangchar}) we 
derive
\be
\chih_{\{n_i\}}(z) = 
\sum_{b_1,b_2,\ldots,b_{N-1}}
\overline{\chi}_{\{b_i\}}^Y(z) \ ,
\ee
where $b_i=0,1$ if $n_i \neq n_{i+1}$ and $b_i=1$ if 
$n_i=n_{i+1}$.

The $q$-factors in the expression for $\ch_{\WZW}^{N,l}(q,z)$
can be understood in the following way. First of all, the 
minimal sequence $n_1,n_2,\ldots,n_N$ that allows placing 
$l$ hooks is obtained from the one for $l-1$ hooks by 
incrementing $l-1$ of the $n_i$ by one, leading to a total 
added energy of $\half l(l-1)$ and a factor $q^{\half l(l-1)}$. 
Secondly, the term with $l$ 
hooks involves a factor
\be
{ 1 \over (q)_l (q)_{N-nl}} \ ,
\label{f3}
\ee
which arises by taking into account a number of
combinatorial factors. For $l=0$ this factor equals
$1/(q)_N$ and arises from summing over the $\{n_i\}$
with given ordering. For an example with non-zero $l$, 
consider the $l=1$ term for $N=n$. It involves a
factor
\be
{1 \over (q)_n} \, (1-q)(1-q^2)\ldots(1-q^{n-1})
\label{f1}
\ee
expressing that the $n$ hooked spinons can no longer
propagate independently and a factor
\be
1+q+q^2+\ldots+q^{n-1} = {1-q^n \over (q)_1} 
\label{f2}
\ee
coming from the fact that there are $n$ distinct
possibilities for placing a single hook (see Appendix~B). 
Together, the factors (\ref{f1}) and (\ref{f2}) produce 
the desired factor (\ref{f3}). For general $N,l$,
the factors (\ref{f3}) can be established using  
induction arguments.

We have thus reduced the general expression for the
$N$-spinon contribution to the CFT partition function
to an expression in terms of the characters of a
conjugate HS model and we can use the explicit results of
section 3.2 to simplify the formulas. Writing
\be
\ch_{\WZW}^N(q,z) = \sum_\lambda z^\lambda \, 
               c_\lambda^{\Lambda,N}(q) \ ,
\ee
where $\Lambda = \Lambda_{N\, \mod \, n}$ and
$\lambda \in P^{(N\, \mod \, n)}$, we obtain
from (\ref{HSchar}) the following result for the
$N$-spinon cut of the $\whsun_1$ string functions
$c^\Lambda_\lambda(q)$
\bea
c_\lambda^{\Lambda,N}(q) &=& 
  q^{\half|\lambda|^2} \sum_{l\geq 0} (-1)^l
  {q^{\half l(l-1)} \over (q)_l} \,
  {1 \over (q)_{N-nl}} M_{\lambda}^{N-nl,n}(q)
\nonu
  &=& 
  q^{\half|\lambda|^2} \sum_{l\geq 0} (-1)^l
  {q^{\half l(l-1)} \over (q)_l} \, 
  {1 \over (q)_{A_1-l} \ldots (q)_{A_n-l}} \ ,
\label{cN}
\eea
where $A_i = {N \over n} - (\lambda,\eps^i)$
(we used that the weights of the representation
${\bf \nb}$ are $-\eps^i$). It follows that 
\be
\Phi_\lambda^{\Lambda,N}(q) =
    \sum_{l\geq 0} (-1)^l
    { q^{\half l(l-1)} \over (q)_l } \,
    { q^{-{(N-nl)(N-nl-n)\over 2n}} \over (q)_{N-nl}}
    K_{( \{A_i-l\}),(1^{N-nl})}(q)
\ee
with Kostka polynomials $K_{\lambda,\mu}(q)$ as
in (\ref{PhiHS}). Finally
\be
\ch_{\WZW}^N(q,z) = q^{-{N^2 \over 2n}} 
  \sum_{\sum_i j_i = N} \, \sum_{l\geq 0} \, 
  (-1)^l {q^{\half l(l-1)} \over (q)_l}
  { q^{\half \sum_i j_i^2} \over \prod_i (q)_{j_i-l}}
  \, z^{- \sum_i j_i \eps^i} \ .
\ee

\newsubsection{Discussion}

The formula (\ref{cN}) allows for the following
interpretation. The $l=0$ contribution to the string 
function $c^{\Lambda,N}_\lambda$, which starts at level
$L_0 = \half|\lambda|^2$ irrespective of $N$, 
describes $N$-spinon states modded out by the homogeneous 
part of the GCR. The inhomogeneous GCR are taken into 
account by the alternating sum over $l\geq 0$. Due to 
this structure, some of the features of the $N$-spinon 
spectrum are not manifest in (\ref{cN}); for example, the
energy $L_0^{{\rm min},N}$ (see (\ref{L0min}))
of the lowest $N$-spinon states is hidden from 
the eye.

In order to get a more transparent formula, we
have tried to trade the alternating sum (over $l$)
for a sum (over a set $l_i$) without alternating 
signs. For this purpose we had to give up the manifest
$su(n)$ symmetry and choose a preferred direction in 
the weight lattice $P$. We then obtained 
\bea
\lefteqn{c_\lambda^{\Lambda,N}(q) =}
\nonu &&
  q^{\half|\lambda|^2} \sum_{l_1,\ldots,l_{n-2}}
  {1 \over \prod (q)_{l_i}} \,
  { q^{A_nl_1+(A_{n-1}-l_1)l_2 + \ldots +
  (A_2-(l_1+\ldots+l_{n-2}))(A_1-(l_1+\ldots+l_{n-2}))} 
  \over
  (q)_{A_n} (q)_{A_{n-1}-l_1} \ldots (q)_{A_2-(l_1+\ldots+l_{n-2})}
  (q)_{A_1-(l_1+\ldots+l_{n-2})} } \ .
\nonu &&
\label{cNbis}
\eea

For $n=2$ (\ref{cNbis}) reduces to \cite{KMM}
\be
c_{\lambda}^{\Lambda,N}(q) = {q^{\quart N^2} \over 
  (q)_{{N \over 2} + {m \over 2}} 
  (q)_{{N \over 2} - {m \over 2}} }
\label{c2}
\ee
for $\lambda = m \Lambda_1$. One clearly sees that the lowest 
$N$-spinon states have energy $\quart N^2$ and have $su(2)$
character $\chi_N(z)$ since
\be
\chi_N(z) = \sum_{-N\leq m \leq N} z^{m \Lambda_1} \ .
\ee
The lowest $N$-spinon states are of the form (\ref{Nsp})
with $n_1=0$, $n_2=1$, $\ldots$, $n_N = N-1$.

For $n>2$ the interpretation of (\ref{cNbis}) is similar 
but somewhat more involved. Putting $n=3$ and choosing
$\lambda=m_1 \Lambda_1+ m_2 \Lambda_2$,
$\Lambda = \Lambda_{2 m_1 + m_2 \, \mod \, 3}$, and
$N=3m_0+2m_1+m_2$, we have 
\be
c_{\lambda}^{\Lambda,N}(q) 
= q^{{N^2 \over 12}} \sum_{l\geq 0} 
  { 1 \over (q)_l } \, 
  {q^{\quart (m_2-m_0+2l)^2} \over 
  (q)_{m_0+m_1+m_2} (q)_{m_0+m_1-l} (q)_{m_0-l}} \ .
\ee
Note that $\lambda$ can also be written as 
$\lambda=-\sum_{i=1}^3 j_i \eps^i$ where $j_1=m_0$, 
$j_2=m_0+m_1$, $j_3=m_0+m_1+m_2$. 
Let us assume that $N$ is even and look at the lowest level 
($L_0={N^2 \over 12}$) contributions to the spectrum. 
These arise from a sum over 
$l=1,2,\ldots,{N \over 2}$. For given $l$, the weights $\lambda$ 
that contribute are selected by $m_2-m_0+2l=0$, $m_0+m_1-l \geq 0$ 
and $m_0-l \geq 0$. Together, these weights constitute the weights 
of the $su(3)$ representation with Dynkin labels $({N \over 2} \; 0)$, 
as expected. This last result is the case $m_2=0$ of the identity 
(we write $\chi_{m_1,m_2}(z)$ for the character of the irreducible
$su(3)$ representation with Dynkin labels $(m_1 \; m_2)$)
\be
\chi_{m_1,m_2}(z) = 
\sum_{l=0}^{m_1} \, \sum_{l'=0}^{m_2}
\sum_{ \scriptsize{ \begin{array}{c} 
 \sum j_i = 2m_1+m_2 \\
 j_1+j_2-j_3=-m_2+2(l+l') \\
 j_1 \geq l \, , \; j_2 \geq l 
 \end{array} }} z^{-\sum_i j_i \eps^i} \ .
\label{charsu3}
\ee
For $n=3$, $N$ odd, this same identity with $m_2=1$ can be used 
to show that the leading $su(3)$ representation, at energy 
${N^2+3 \over 12}$, is indeed the one with Dynkin labels 
$({N-1 \over 2}\; 1)$. For general $n$, a similar analysis is 
easily performed.

There is a good reason for the fact that the $n>2$ 
character formulas are quite a bit more complicated than the 
ones for $n=2$. That reason is the way the Generalized
Pauli Principle works out for the lowest $N$-spinon
excitations over the ground state. For $n=2$ the Generalized
Pauli Principle tells us to symmetrize over $N$ `adjacent' 
spinons, leading to the simple character formula (\ref{c2}). 
A naive generalization to $n>2$ would have led to a 
lowest $N$-spinon state which is a  symmetric combination 
of $N$ adjacent spinons, but this is not the correct result. 
Instead, we have seen that the lowest state with, say, $N=p(n-1)$ 
spinons of $su(n)$, is obtained by {\em (i)}\ antisymmetrizing over 
groups of $n-1$ spinons with equal mode indices $n_i$, 
giving a representation ${\bf n}$ for each group, and {\em (ii)}\ 
symmetrizing over $p$ adjacent groups, giving a representation 
with Dynkin labels $(p \, 0 \, \ldots)$. The corresponding 
character formulas can be viewed as affinizations of 
identities of the type (\ref{charsu3}).

\newpage

\newsection{Alternative formulations}

\vskip -2mm

\newsubsection{More spinons}

\noindent

It will be clear that in our analysis so far a
very special role was played by products
\be
 \ldots (\phi^{\alpha_{i+k}}_{x_{i+k}-n} \ldots 
 \phi^{\alpha_{i+1}}_{x_{i+1}-n}) \ldots 
\ee
($x_j=-{n-(2j-1) \over 2n}$) of $k$ consecutive spinon modes with 
equal index $n$. The Generalized Pauli Principle of section 4.3
tells us that in this situation the $su(n)$ indices 
$\alpha_{i+1},\ldots,\alpha_{i+k}$ should be anti-symmetrized, so 
that the `block' of $k$ bound spinons carries $su(n)$ representation 
$L(\Lambda_{n-k})$. It is then natural to consider a formulation
that has independent spinon fields $\phi^{(j)}$, $j=1,2,\ldots,n-1$, 
transforming in $L(\Lambda_j)$ of $su(n)$. Of course, these spinons
simply correspond to the $n-1$ independent primary fields of the
$\whSUn_1$ WZW theory. (Such a description was also suggested in 
\cite{Re} by analysis of the $SU(n)$ Heisenberg chain in the infinite 
chain limit.)

We can thus take a point of view where the chiral Hilbert space of 
a WZW theory is built by acting with the modes of a set of $n-1$ 
independent spinon fields, modded out by Generalized Commutation 
Relations and/or null fields. In this section we focus on 
$su(3)$ to illustrate such an approach. The $su(n)$ case
will be discussed in \cite{BS2}.

In section 5.2 we shall reformulate the approach based on 
the yangian $Y(sl_3)$ in terms of spinons $\phi^{(1)}$ and 
$\phi^{(2)}$ with mode indices that are related to parameters 
$m^1_1,\ldots,m^1_{M_1}$ and $m^2_1,\ldots,m^2_{M_2}$,
which are the zero's of Drinfel'd polynomials $P_1(u)$ and 
$P_2(u)$.

Clearly, $\phi^{(1)}$ and $\phi^{(2)}$, which transform in the
${\bf 3}$ and ${\bf \threeb}$ of $su(3)$, respectively, are
related by a discrete symmetry (the conjugation symmetry of 
$su(3)$). However, the yangian generators do not respect this 
symmetry and it follows that in the yangian formulation the 
spinons $\phi^{(1)}$ and $\phi^{(2)}$ are not on equal footing. 
In section 5.3 we present alternative character formulas that 
do respect the symmetry between $\phi^{(1)}$ and $\phi^{(2)}$, 
meaning that conjugate pairs of physical states are labeled by 
conjugate sets of parameters. Depending on the application one
has in mind, one may prefer either the `yangian formulation'
or the `symmetric formulation'.

\newsubsection{Yangian irreducible representations, III}

In section 4, we characterized a motif-related yangian representation 
in the spectrum of the $\whSUn_1$ CFT by a set of mode indices 
$\{n_i\}$. Following our observation in 5.1, we may trade pairs of 
equal indices, $n_i=n_{i+1}$ for parameters $m^1_j$ and single $n_i$ 
for parameters $m^2_j$. We then obtain a parametrization of motifs
in terms of parameters $\{m^1_j\}$, $\{m^2_j\}$, which have the 
interpretation of zero's of Drinfel'd polynomials $P_1(u)$ and $P_2(u)$. 

The $m^1_j$ are integer while the $m^2_j$ are half-odd-integer. 
Placing them in increasing order, they should satisfy the following. 
The smallest is of the form $m^1_1=1+3n$ or $m^2_1={3 \over 2}+3n$
with $n$ a non-negative integer. The increment between adjacent 
$m^1_j$ is of the form $1+3n$, between adjacent $m^1_j$ and 
$m^2_j$ it is ${3 \over 2}+3n$, and between adjacent $m^2_j$ it is 
$2+3n$. 

For given $\{m^1_j\}$, $\{m^2_j\}$, the corresponding motif is
reconstructed as follows. Every $m^1_j$ is replaced by `()',
every $m^2_j$ by `(1)'. If two $m^i_j$ are separated by the
minimal increment plus $3n$ we place $n$ times `(11)' between
the corresponding motifs, and we place `$(11)(11)\ldots$' to
the right of the largest $m^i_j$. Finally, we replace the initial 
`(' and all `)(' by `0'. This connection was first explained in 
\cite{BGHP}. 

Translating the yangian character formula (\ref{yangchar2}) to 
this new formulation, we find the following. We first define a 
character
\be
  \chih_{\{m^1_j\},\{m^2_j\}}(z) \ ,
\label{chihm1m2}
\ee
which is obtained by writing a factor $\chi_{l,0}(z)$ for a
group of $l$ adjacent $m^1_j$ with minimal increment 1 and
a factor $\chi_{0,1}(z)$ for every $m^2_j$. If the increment
between adjacent $m_1^j$ and $m_2^j$ is minimal, \ie,
equal to ${3 \over 2}$, we allow a contraction, meaning that 
we replace a product 
$\chi_{0,1}(z)\chi_{l,0}(z)$ (or $\chi_{l,0}(z)\chi_{0,1}(z)$) 
by the trace $\chi_{l-1,0}(z)$. The yangian character is then 
written as
\be
  \chi^Y_{\{m^1_j\},\{m^2_j\}}(z) = \sum_{l\geq 0}(-1)^l
    \chih^{(l)}_{\{m^1_j\},\{m^2_j\}}(z) \ ,
\label{yangchar3}
\ee
where $\chih^{(l)}_{\{m^1_j\},\{m^2_j\}}(z)$ is the character
(\ref{chihm1m2}) with $l$ contractions. Let us stress that the
character formula (\ref{yangchar3}) is only applicable to
those Drinfel'd polynomials that occur in the yangian 
decomposition of the HS and WZW modules and does not describe
the most general irreducible representations of $Y(sl_3)$.

As an example of (\ref{yangchar3}), consider $m^1_1=1$, 
$m^2_1={5 \over 2}$, which corresponds to the motif 
`$001011011\ldots$'. We have
\be
  \chi^Y_{\{1\}, \{ {5 \over 2}\} }(z) = 
  \chi_{1,0}(z)\chi_{0,1}(z) - \chi_{0,0}(z)
\ee
in agreement with (\ref{char01}). Note that a single
character $\chi_{l,0}$ may be contracted twice, as in
\be
  \chi^Y_{ \{3,4\} , \{{3 \over2},{11 \over 2}\} }(z) = 
  (\chi_{0,1}\chi_{2,0}\chi_{0,1} - \chi_{1,0}\chi_{0,1}
   - \chi_{0,1}\chi_{1,0} + \chi_{0,0})(z) \ ,
\ee
which is the character for the motif `$010001011011\ldots$'.

The energy of the states labeled by $\{m^1_j\}$, $\{m^2_j\}$ 
comes out as
\be
L_0 = -{2M_1+M_2 \over 6}
  + {1 \over 3} \sum_j (2m^1_j + m^2_j) \ ,
\ee
and we can write the following character formula for the states
with `spinon-numbers' $M_1$ and $M_2$
\be
\chh_{\WZW}^{M_1,M_2}(q,z) = \sum_{ 
  \scriptsize{ \begin{array}{c}
  m^1_1, \ldots, m^1_{M_1} \\
  m^2_1, \ldots, m^2_{M_2} \end{array} }}
  q^{-{2M_1+M_2 \over 6}
  + {1 \over 3} \sum_j (2 m^1_j + m^2_j)} \,
  \chi^Y_{ \{ m^1_j \} , \{m^2_j\} }(z) \ .
\label{chM1M2}  
\ee
By construction, this character has the property
\be
\ch_{\WZW}^N(q,z) = \sum_{ 2M_1+M_2 =N} 
              \chh_{\WZW}^{M_1,M_2}(q,z) \ .
\ee
We would like to repeat once again that the character 
(\ref{chM1M2}) is based on a decomposition of the spectrum in 
terms of yangian multiplets, and does thereby not treat the 
conjugate spinons $\phi^{(1)}$ and $\phi^{(2)}$ on equal 
footing.

\newsubsection{More character formulas}

We already mentioned that it is natural to look for a 
formulation which respects the conjugation symmetry between 
the fields $\phi^{(1)}$ and $\phi^{(2)}$. Note that
this means that we give up the connection with the yangian
symmetry and are no longer working with eigenstates
of operators such as $H_2$ (in the paper \cite{AN},
this point was not fully appreciated).

For $\widehat{su(3)}_1$, we have found a character formula with 
the expected structure. To write it, we first define a character 
$\ch_{\rm F}^{N_1,N_2}(q,z)$, 
which is to be viewed as the character of a `Fock space' of a 
total of six (\ie, the components of $\phi^{(1)}$
and $\phi^{(2)}$) sets of spinon modes
\be
\ch_{\rm F}^{N_1,N_2}(q,z) = 
  \sum_{\sum_i j^1_i = N_1, \, \sum_i j^2_i = N_2} 
  { q^{\third\left( N_1^2 + N_2^2 + N_1N_2 \right)} \over
  \prod_i (q)_{j_i^1} (q)_{j_i^2} }
  \, z^{\sum_i (j^1_i - j^2_i) \eps^i} \ .
\label{charF}
\ee
This character is an affinization of 
\be
\chih_{m_1,m_2}(z) = 
 \sum_{\sum_i j^1_i = m_1, \, \sum_i j^2_i = m_2}
 z^{\sum_i (j^1_i - j^2_i) \eps^i} \ ,
\ee
which is related to the irreducible $su(3)$ character via
\be
\chi_{m_1,m_2}(z) 
  = \chih_{m_1,m_2}(z) - \chih_{m_1-1,m_2-1}(z) \ .
\ee

The irreducible affine characters should be affinizations
of the irreducible $su(3)$ characters and we therefore
expect an alternating sum to make its appearance. In close 
analogy with (\ref{WZWN}) we define 
\be
\ch_{\WZW}^{N_1,N_2}(q,z) =
\sum_{l\geq 0} 
(-1)^l {q^{\half l(l-1)} \over (q)_l} \,
q^{l(N_1+N_2-l)} \, \ch_{\rm F}^{N_1-l,N_2-l}(q,z) \ .
\label{charn1n2}
\ee
It can be proved that 
\be
\ch_{\WZW}^N(q,z) = \sum_{ 2N_1+N_2 =N} 
                    \ch_{\WZW}^{N_1,N_2}(q,z) 
\ee
so that
\be
\ch_{\Lh(\Lambda_{3-k})}(q,z) 
  = \sum_{2N_1 + N_2 \equiv k \, \mod \, 3}
  \ch_{\WZW}^{N_1,N_2}(q,z) \ .
\ee
We have thus succeeded in breaking down $N$-spinon characters 
into $(N_1,N_2)$-spinon characters with manifest conjugation 
symmetry.

In a follow-up paper \cite{BS2}, we shall extend the discussion 
of this section to general $su(n)$ and further investigate 
character formulas such as (\ref{charn1n2}). 

\newpage

\newsection{Conclusions}

\vskip 2mm

The results presented in this paper should be viewed as a first 
step towards a full-fledged multi-spinon formulation of these 
(and other) CFT's. As a next step, one would like to construct 
explicit multi-spinon bases for the Hilbert space, and deal with 
correlation functions and form factors. For the case 
$\widehat{SU(2)}_1$, part of such a program has already been 
completed \cite{BPS,BLS1,BLS1a}.

At the mathematical level, there are several interesting aspects
to the results presented in this paper. We have given a number of 
equivalent formulae for the characters of a large class of 
finite-dimensional irreducible representations of the yangian 
$Y(sl_n)$. Thus far, these characters were not known.  
Expressing them in terms of Schur polynomials we
find a new interesting class of symmetric functions (see \cite{BS2}).
We believe that a proper understanding of the origin of these 
symmetric functions will ultimately lead to an understanding of all
$Y(sl_n)$ irreducible finite-dimensional representations.

We have also seen an intricate connection between $\widehat{SU(n)}_1$
CFT's and the Haldane-Shastry model. On the one hand, the 
$\widehat{SU(n)}_1$ theory can be obtained as the $L \to \infty$ limit
of an $L$-site HS model, while on the other hand, the $N$-spinon cuts
of the affine characters can be expressed as an alternating sum
over characters of the $(N-nl)$-site ($l\geq0$) conjugate HS model.
This `duality' deserves to be better understood.

Apart from the issues that remain for the $\whSUn_k$ WZW models, it 
is interesting to consider other groups. Of particular interest are 
the $\widehat{SO(n)}_1$ WZW models, which are related to $n$ Majorana 
fermions via non-abelian bosonization and are strongly expected to 
show up in condensed matter systems. The challenge here is to give a 
formulation in terms of fundamental quasi-particles that are spinors 
(of conformal dimension $n/16$) of $so(n)$. The case $\widehat{so(3)}_1$ 
is formally the same as $\widehat{su(2)}_2$ and is thus covered by the 
analysis of \cite{BLS2}. Similarly, $\widehat{so(6)}_1$ is the same 
as $\widehat{su(4)}_1$, which was treated in this paper. Full results 
for general $n$ will be given elsewhere.

There are other examples (other than WZW models) of CFT's that 
describe quasi-particles of fractional statistics (in a sense,
{\em any}\ rational CFT has an interpretation of this type). 
Perhaps the simplest among these are $c=1$ theories at radius 
$R^2 = p/q$. For $p=2$, $q=1$ this is nothing else than the 
$\widehat{SU(2)}_1$
WZW theory but for more general $p,q$ these theories do not
have Lie algebra symmetries. Choosing $p=2m+1$, $q=1$ gives a 
theory describing one-component edge excitations of a Fractional 
Quantum Hall (FQH) sample of filling fraction $1/(2m+1)$ (they 
are the case $n=1$ of the $n$-component edge theories mentioned 
in the introduction). A `multi-spinon formulation' \cite{Is,Sc2} 
of these CFT's directly corresponds to a description of the FQH 
edge dynamics in terms of the fundamental edge degrees of freedom 
of charge $e/(2m+1)$ and statistics $\theta=\pi/(2m+1)$.

\vskip 1cm
\noindent {\em Acknowledgements.}\ 
PB would like to thank the Institute for Theoretical Physics at 
the University of Amsterdam for hospitality. KS acknowledges the 
ITP at the University of Adelaide for hospitality and thanks the 
organizers of the workshop on `Infinite Analysis' at the International 
Institute for Advanced Study in Kyoto/Nara for the invitation
to participate. Part of this work was done in these stimulating 
environments. PB is supported by an 
Australian Research Council Fellowship and Small Grant while KS is 
supported in part by the Foundation FOM of the Netherlands.

\newpage

\appendix

\newsection{Illustration: $\widehat{SU(3)}_1$}

We illustrate the spinon structure of the 
$\widehat{SU(3)}_1$ CFT in Table 2 below.

\vskip 10mm

\begin{center}
\begin{tabular}{|cccllc|}
\hline
\str $L_0$ & $H_2$ & $N$ & $\{n_i\}$ & 
    motif & $SU(3)$ content \\
\hline
\str
0 & 0 & 0 &  & 
    $011011011 \ldots$ & ${\bf 1}$ \\
\str
${1 \over 3}$ & ${1 \over 9}$ & 1 & 0 & 
    $01011011 \ldots$ & ${\bf \threeb}$ \\
\str
${1 \over 3}$ & $-{1 \over 9}$ & 2 & 0,0 & 
    $0011011011 \ldots$ & ${\bf 3}$ \\ 
\str
1 & 1 & 3 & 0,0,1 & 
    $001011011 \ldots$ & ${\bf 8}$ \\
\str
${4 \over 3}$ & ${46 \over 9}$ & 1 & 1 & 
    $01101011\ldots$ & ${\bf \threeb}$\\
\strr
${4 \over 3}$ & ${10 \over 9}$ & 4 & 0,0,1,1 & 
    $00011011\ldots$ & ${\bf 6}$\\
\str
${4 \over 3}$ & ${26 \over 9}$ & 2 & 0,1 & 
    $0101011011\ldots$ & ${\bf 3} + {\bf \sixb}$\\
\str
2 & 4 & 3 & 0,1,1 & 
    $010011011 \ldots$ & ${\bf 8}$ \\
\strr
2 & 8 & 3 & 0,0,2 & 
    $001101011 \ldots$ & ${\bf 1}+{\bf 8}$ \\
\str
${7 \over 3}$ & ${155 \over 9}$ & 1 & 2 & 
    $01101101011 \ldots$ & ${\bf \threeb}$ \\ 
\strr
${7 \over 3}$ & ${55 \over 9}$ & 4 & 0,0,1,2 & 
    $00101011011 \ldots$ & ${\bf \threeb} + {\bf 6} + {\bf \fifteenb}$ \\ 
\str
${7 \over 3}$ & ${71 \over 9}$ & 2 & 1,1 & 
    $0110011011 \ldots$ & ${\bf 3}$ \\ 
\strr
${7 \over 3}$ & $ {107\over 9}$ & 2 & 0,2 & 
    $0101101011 \ldots$ & ${\bf 3} + {\bf \sixb}$ \\ 
\strr
${7 \over 3}$ & ${35 \over 9}$ & 5 & 0,0,1,1,2 & 
    $0001011011 \ldots$ & ${\bf 15}$  \\
\str
3 & 21 & 3 & 0,0,3 &
    $001101101011 \ldots$ & ${\bf 1}+{\bf 8}$ \\ 
\strr
3 & 11 & 3 & 0,1,2 &
    $010101011011 \ldots$ & ${\bf 1}+{\bf 8}_2+{\bf \tenb}$\\ 
\strr
3 & 5 & 6 & 0,0,1,1,2,2 &
    $000011011 \ldots$ & ${\bf 10}$ \\ 
\hline
\end{tabular}
\end{center}
\vskip 5mm
\begin{quote}
Table 2: Spinon structure of the $L_0\leq 3$
states in the spectrum of the $\widehat{SU(3)}_1$ WZW model. 
The eigenvalues of $H_2$ are as in (\ref{htwo}) and the $su(3)$ 
structure is as explained in section 4.3. Note the asymmetry in 
the description of conjugate states in the spectrum.
\end{quote}

\newpage

\newsection{Hook rules}

We present the rules for placing hooks in the construction 
of section 4.3 of the character (\ref{yangchar2}) of an 
irreducible representation of $Y(sl_n)$. 
Starting point is a mode sequence $n_1,n_2,\ldots,n_N$.
The rules for placing a single hook are
\begin{enumerate}
\item
A hook 
\be
\underhook{n_i \ldots n_{i+n-1}} 
\ee
may be placed in the following situations
\bea &&
n_i=n_{i+1}=\ldots =n_{i+n-1} \ ,
\nonu &&
n_i= \ldots = n_{i+k-1} \isone n_{i+k} = \ldots
 = n_{i+n-1}\ , \qquad k=1,\ldots,n-1 \ . 
\eea
\item
The hook
\be
n_{i-1} \, \underhook{n_i \ldots n_{i+n-1}}  
\ee
is allowed if $\underhook{n_i \ldots n_{i+n-1}}$ is
allowed and $n_{i-1} < n_i$.
\item
The hook
\be
\underhook{n_i \ldots n_{i+n-1}} \, n_{i+n}  
\ee
is allowed if $\underhook{n_i \ldots n_{i+n-1}}$ is
allowed and $n_{i+n-1} < n_{i+n}$, with the following
exceptions. The hook 
\be
\underhook{n_i \ldots n_{i+n-1}} \, n_{i+n} \ldots n_{i+2n-1} 
\ee
is {\em not}\ allowed if 
$n_i = \ldots = n_{i+n-1} \isone n_{i+n} = \ldots =n_{i+2n-1}$, 
while (for $n>2$) the hook
\be
\underhook{n_i \ldots n_{i+n-1}} \, n_{i+n} \ldots 
\ee
{\em is}\ allowed if $n_{i+n-1}=n_{i+n}$ and the total
number of $n_j$ equal to $n_{i+n}$ is precisely $n-1$.
\end{enumerate}
To determine if additional hooks can be placed {\em to the 
right of}\ existing hooks, one removes all hooked $n_i$ from 
the sequence and then applies the above rules to the 
remaining $n_i$.

\newsection{Decomposition of motifs}

In this appendix we clarify further the $su(n)$ content of 
irreducible yangian representations. Let us consider a motif 
given by a sequence of `0' and `1'.

In \cite{HHTBP} it was proposed that the $su(n)$ content 
of a motif can be obtained as a (free) tensor product of 
a number of component motifs. For $su(3)$ this decomposition 
is obtained by {\em (i)}\ replacing the initial and final `0'
by `(' and `)', respectively, {\em (ii)}\ replacing all `0110' 
by `)(11)(' and {\em (iii)}\ replacing all `101' by 1)(1'. 
For example
\be
010110 \; \rightarrow (1)(11) \ ,
\ee
giving $su(3)$ content ${\bf \threeb}$.

The same motif for $su(5)$ has content ${\bf 75} \oplus 
{\bf 24}$, which is different from the product ${\bf 10}
\otimes {\bf \tenb}$ of the motifs $(1)$ and $(11)$
(hence the name `incomplete multiplets' in \cite{Sc1}).
This clearly indicates that the above rule needs to be 
changed for general $n$. The correct rule that replaces
{\em (ii)}\ and {\em (iii)}\ is to replace all `0' that are 
adjacent (on either side) to a total of at least $n-1$ `1' by
`)('. (This is most easily proved by using the description of 
section 4.3.) According to this rule, the motif `010110' is 
indecomposable for $n\geq 5$.

Note that our prescriptions in sections 3.1, 4.3 and
5.2 do not make use of this decomposition rule.


\newpage

\frenchspacing
\baselineskip=16pt
\begin{thebibliography}{11}

\bibitem{Ha2}
  F.D.M. Haldane, Phys. Rev. Lett. {\bf 66} (1991) 1529.
\bibitem{FLS}
  P. Fendley, A.W.W. Ludwig and H. Saleur,
  Phys. Rev. Lett. {\bf 74} (1995) 3005, ({\tt cond-mat/9408068}).
\bibitem{BPS}
  D. Bernard, V. Pasquier and D. Serban, 
  Nucl. Phys. {\bf B428} (1994) 612, {\tt (hep\-th/9404050)}.
\bibitem{BLS1}
  P. Bouwknegt, A.W.W. Ludwig and K. Schoutens,
  Phys. Lett. {\bf 338B} (1994) 448, {\tt (hep-th/9406020)}.
\bibitem{BLS1a}
  P. Bouwknegt, A.W.W. Ludwig and K. Schoutens,
  in Proc. of the 1994 Trieste Summer School on
  ``High Energy Physics and Cosmology'' Trieste, 
  July 1994, ({\tt (hep-th/949412199)}.
\bibitem{FS}
  B.L.~Feigin and A.V.~Stoyanovsky, {\it Quasi-particle models for the
  representations of Lie algebras and geometry of flag manifolds},
  ({\tt hep-th/9308079}).
\bibitem{Ge}
  G.~Georgiev, {\it Combinatorial constructions of modules for 
  infinite-dimensional Lie algebras, I.\ Principal subspace},
  ({\tt hep-th/9412054}); {\it ibid.}, 
  {\it II.\ Parafermionic space}, ({\tt q-alg/9504024}).
\bibitem{BS2}
  P. Bouwknegt, K. Schoutens, in preparation.
\bibitem{BLS2}
  P. Bouwknegt, A.W.W Ludwig and K. Schoutens,
  Phys. Lett. {\bf 359B} (1995) 304, {\tt (hep-th/9412108)}.
\bibitem{NYa}
  A.~Nakayashiki and Y.~Yamada, Comm.\ Math.\ Phys.\ 
  {\bf 178} (1996) 179, ({\tt hep\-th/9504052}).
\bibitem{ANOT}
  T.~Arakawa, T.~Nakanishi, K.~Oshima and A.~Tsuchiya, 
  {\it Spectral decomposition of path space in solvable lattice model},
  ({\tt q-alg/9507025}).
\bibitem{Ka}
  V.G.~Kac, {\it Infinite dimensional Lie algebras}, (Cambridge 
  University Press, Cambridge, 1985).
\bibitem{Dr}
  V.G. Drinfel'd, Sov. Math. Dokl. {\bf 32} (1985)
  254; in Proceedings of the International Congress of 
  Mathematicians, Berkeley, California (1986).
\bibitem{CPa}
  V.~Chari and A.~Pressley, {\it A guide to Quantum Groups}, 
  (Cambridge University Press, Cambridge, 1994).
\bibitem{Sc1}
  K. Schoutens,
  Phys. Lett. {\bf 331B} (1994) 335, {\tt (hep-th/9401154)}.
\bibitem{BS1}
  P. Bouwknegt, K. Schoutens, Phys. Rep. 
  {\bf 223} (1993) 183, {\tt (hep-th/9210010)}.
\bibitem{Ha1}
  F.D.M. Haldane, Phys. Rev. Lett. {\bf 60} (1988) 635.
\bibitem{Sh}
  B. Sriram Shastry, Phys. Rev. Lett. {\bf 60} (1988) 639.
\bibitem{HHTBP}
  F.D.M. Haldane, Z.N.C. Ha, J.C. Talstra, D. Bernard
  and V. Pasquier, Phys. Rev. Lett. {\bf 69} (1992) 2021.
\bibitem{HT}
  J.C. Talstra and F.D.M. Haldane, {\it Integrals of motion
  of the Haldane Shastry model}, {\tt (cond-mat/9411065)}.
\bibitem{BGHP} 
  D.~Bernard, M.~Gaudin, F.D.M.~Haldane and V.~Pasquier,
  J.\ Phys. {\bf A26} (1993) 5219, ({\tt hep-th/9301084}).
\bibitem{HH}
  F.D.M. Haldane, Z.N.C. Ha, Phys. Rev. {\bf B47} (1993)
  12459.
\bibitem{CPb}
  V.~Chari and A.~Pressley, {\it Yangians: their representations and 
  characters}, ({\tt q-alg/9508004}).
\bibitem{Me}
  E. Melzer, Lett. Math. Phys. {\bf 31} (1994) 233,
  {\tt (hep-th/9312043)}.
\bibitem{Ki}
  A.N.~Kirillov, {\it Dilogarithm identities}, 
  {\tt (hep\-th/9408113)}.
\bibitem{NY}
  A.~Nakayashiki and Y.~Yamada, {\it Kostka polynomials and 
  energy functions in solvable lattice models}, 
  ({\tt q-alg/9512027}).
\bibitem{AN}
  C. Ahn and S. Nam, {\it Yangian Symmetries in the $SU(N)_1$
  WZW Model and the Calogero-Sutherland Model},
  preprint SNUTP/95-113, {\tt (hep-th/9510242)}.
\bibitem{Ha3}
  F.D.M. Haldane, Phys. Rev. Lett. {\bf 67} (1991) 937.
\bibitem{KMM}
  R. Kedem, B. McCoy and E. Melzer, in Proc. of the 
  conference SMQFT, USC 1994, P.~Bouwknegt et al, eds
  (World Scientific, 1995), {\tt (hep-th/9304056)}.  
\bibitem{Re}
  N.~Reshetikhin, J.\ Phys.\ {\bf A24} (1991) 3299.
\bibitem{Is}
  S. Iso, Nucl. Phys. {\bf B443} (1995) 581,
  ({\tt hep-th/9411051}). 
\bibitem{Sc2}
  K. Schoutens, unpublished.
\end{thebibliography}
\end{document}